\documentstyle[12pt]{article}
\begin{document}
\def\om{\omega}
\def\g{\gamma}
\def\omt{\tilde{\omega}}
\def\ti{\tilde}
\def\o{\Omega}
\def\t{T^*M}
\def\vt{\tilde{v}}
\def\ot{\tilde{\Omega}}
\def\otwo{\omt \wedge \om}
\def\owot{\om \wedge \omt}
\def\w{\wedge}
\def\mt{\tilde{M}}
\def\T{{\rm Tr}}
\def\om{\omega}
\def\omt{\tilde{\omega}}
\def\ss{\subset}
\def\bc{{\bf C}}
\def\th{\theta}
\def\k{\kappa}
\def\bk{\bar\kappa}
\def\bx{\bar x}

\def\om{\omega}
\def\omt{\tilde{\omega}}
\def\ti{\tilde}
\def\o{\Omega}
\def\t{T^*M}
\def\vt{\tilde{v}}
\def\ot{\tilde{\Omega}}
\def\otwo{\omt \wedge \om}
\def\owot{\om \wedge \omt}
\def\w{\wedge}
\def\mt{\tilde{M}}
\def\g{\gamma}
\def\om{\omega}
\def\omt{\tilde{\omega}}
\def\ss{\subset}
\def\tpm{T_{P} ^* M}
\def\al{\alpha}
\def\alt{\tilde{\alpha}}
\def\la{\langle}
\def\ra{\rangle}
\def\inop{{\int}^{P}_{P_{0}}{\om}}
\def\th{\theta}
\def\lm{\lambda}
\def\tht{\tilde{\theta}}
\def\inox{{\int}^{X}{\om}}
\def\inotx{{\int}^{X}{\omt}}
\def\st{\tilde{S}}
\def\ls{\lambda_{\sigma}}
\def\p{{\bf{p}}}
\def\pb{{\p}_{b}(t,u)}
\def\pbm{{\p}_{b}}
\def\d{\partial}
\def\d+{\partial_+}
\def\d-{\partial_-}
\def\pat{\partial_{\tau}}
\def\pas{\partial_{\sigma}}
\def\dpm{\partial_{\pm}}
\def\l2{\Lambda^2}
\def\be{\begin{equation}}
\def\ee{\end{equation}}
\def\bea{\begin{eqnarray}}
\def\eea{\end{eqnarray}}
\def\ej{{\bf E}}
\def\ed{{\bf E}^\perp}
\def\si{\sigma}
\def\cg{{\cal G}}
\def\cgt{\ti{\cal G}}
\def\cd{{\cal D}}

\def\d{\partial}
\def\dz{\partial_z}
\def\dbz{\partial_{\bar z}}

\def\be{\begin{equation}}
\def\ee{\end{equation}}
\def\bea{\begin{eqnarray}}
\def\eea{\end{eqnarray}}

\def\si{\sigma}

\def\bz{\bar{z}}
\def\e{\varepsilon}
\def\b{\beta}
\begin{titlepage}

\rightline{IHES/P/97/58}
\rightline{hep-th/9707194}

\vspace{3cm}
\begin{center}
{\Large \bf Poisson-Lie T-duality and (1,1) supersymmetry}\\
[50pt]{\small
{\bf C. Klim\v{c}\'{\i}k}\\ IHES, 
91440 Bures-sur-Yvette, France
\\}
\vspace{2cm}
\begin{abstract}
A duality invariant action for (1,1) supersymmetric extension of Poisson-Lie
dualizable $\sigma$-models is constructed.

\end{abstract}
\end{center}
\vskip 1.5cm

\end{titlepage}

1. The Poisson--Lie (PL) $T$-duality \cite{KS2} is a generalization of the 
traditional non-Abelian $T$-duality \cite{OQ}--\cite{GR} and it  is proved
to enjoy 
\cite{KS2},
\cite{AKT}--\cite{KS8}, at least at the classical level,  most of the
structural features of the traditional Abelian $T$-duality 
\cite{SS} and \cite{Busch}.  

Perhaps the last remaining problem to be solved in order to complete 
the classical 
story consists in construction of the supersymmetric (SUSY) extension of the
PL duality. The first step in this direction was undertaken by Sfetsos in 
\cite{Sf} where he noted that 
the (1,1) SUSY extensions (including the spectator
fields) of the PL dual pair 
of $\si$-models have equivalent equations of motion.
In other words, a solution 
of the field equations of one model can be explicitly
mapped to a solution of the dual model. In \cite{Sf} and later in
a series of papers by Parkhomenko \cite{Par} also the case of extended
supersymmetry was  treated, however, the PL duality was always established
only in the "on shell" way described above. After experience with the purely
bosonic case, it is plausible that an "off shell" equivalence of the SUSY
models can be also established which would amount to showing that the map
sending a solution of the one model to the solution of the dual model is
actually a symplectomorhism (or a canonical transformation) between the
phase spaces of the models. In the purely bosonic case, the fact that the
duality map is the symplectomorphism was established in \cite {KS2} by means
of constructing the generating functional of the canonical transformation.
Later in \cite {KS3}, we have given a manifest evidence of the off shell
equivalence of the (purely bosonic) models: we parametrized the common
phase space of the models in a duality invariant way (as a coadjoint orbit
of the loop group of the underlying Drinfeld double) and found  the first-order
duality invariant Hamiltonian action reflecting simultaneously the dynamics
of both models in the dual pair. In the SUSY case, such a duality invariant 
description of the phase space and a duality invariant action are missing;
the goal of this paper is to fill this gap.

The main trouble in addressing the problem consists in the fact that in the
first-order Hamiltonian description the world-sheet supersymmetry cannot be 
manifest. As the matter of the fact, already in the purely bosonic case
one misses the explicit world-sheet Poincar\'e symmetry in the first-order
action \cite {KS3}. However, we were able to establish (with P. \v Severa) 
that the
symplectic structure on the phase space is just the Kirillov-Kostant symplectic
structure on the coadjoint orbit of the loop group and it was sufficient
only to calculate the Hamiltonian which would give the dynamics of the PL
dual pair of $\si$-models. The Poincar\'e symmetry of the theory  then somehow
miraculously followed in the
second-order $\si$-model description.

In what follows, we will work out  the  supersymmetric
case  in the
same spirit.
 Namely, we will find the duality invariant parametrization of the
 configurations of the model and  an explicit formula
for the  duality invariant action. We shall first do
the (1,0) case (which is  itself interesting as it
 is relevant for the description of
 the  heterotic
 string) where we perform a detailed calculation  that illustrates the
 method of obtaining the duality invariant action from the second order
 $\si$-model description. Then we shall use the method to solve the
 (1,1) case.

2. 
For the description of the PL $T$-duality, we need the
crucial concept
of the Drinfeld double,  which is simply  a  Lie group $D$ such that
its Lie algebra $\cd$ (viewed as a vector space) 
 can be decomposed as the direct sum  of two  subalgebras, $\cg$ and $\cgt$, 
maximally isotropic with 
respect to a non-degenerate invariant bilinear form on $\cd$ \cite{D}.
It is often convenient to identify the dual linear space to $\cg$ ($\cgt$)
with $\cgt$ ($\cg$) via this bilinear form.

From the space-time point of view, we have identified 
 the targets of the mutually 
dual $\sigma$-models
with the group manifolds  $G$ and $\ti G$ \cite{KS2}, corresponding to the Lie 
algebras $\cg$ and $\cgt$.

Consider now the (1,0) SUSY extension of the PL dualizable $\sigma$-model
on the target $G$ (cf. \cite{KS2,AKT,KS3}):

\be S={1\over 2}\int d\si d\tau d\theta 
\{\bar\Lambda^a(R_{ab}+\Pi(G)_{ab})\lambda^b + 
\bar\Lambda^a (\partial_-G G^{-1})_a
-(D_+ G G^{-1})_a\lambda^a\}.\ee
Here  the (cylindrical)
world-sheet is parametrized by $\tau\in[-\infty,\infty], \si\in[0,2\pi]$ and
\be \partial_{\pm}=\pat\pm\pas.\ee
The Grassmann variable
$\theta$ is the superpartner of $\xi^+={1\over 2}(\tau+\si)$, hence
\be D_+=\partial_{\theta}+\theta\partial_+.\ee
The superfield $G$ is given in components as
\be G=g+\theta \gamma g, \quad g\in G, \quad \g\in\cg;\ee
$\g$ is the Majorana-Weyl fermion.
The auxiliary superfields $\bar\Lambda^a$
and $\lambda^a$ can be written also as expansions in $\th$ with
generic coefficients; note, however, 
that $\Lambda$ is an 
odd superfield while
$\lambda$ is an even one and they both are not group valued. The components of
the currents $dG G^{-1}$ are taken in a basis $T^a$ of $\cg$.
 Of course, the Gaussian integration over the
auxiliary superfields can be performed, yielding, perhaps, a more standard
way of writing the $2d$ $\si$-model action on a group target.

Finishing the explanations of the symbols in (1), we note that $R_{ab}$ is an
arbitrary non-degenerate constant matrix and $\Pi(G)_{ab}$ is the bivector
field on the group manifold $G$ that gives the famous Poisson-Lie bracket
 on $G$
\cite{D}. For our purposes, it is convenient to describe $\Pi(G)$ more
explicitly; consider matrices $a(g),b(g)$ and $d(g)$ defined as 

\be g^{-1}T^a g=a(g)^a_{~b} T^b, \quad g^{-1}\ti T_a g =
b(g)_{ab}T^b +d(g)_a^{~b} \ti T_b,\ee
where $T^a(\ti T_a)$ are generators of the Lie algebra $\cg(\cgt)\ss\cd$.
Now 
\be \Pi(g) =b(g)a(g)^{-1}.\ee
Note also a few useful properties of the matrices $a,b$ and $d$:
\be d^t(g)=a(g^{-1}),\quad b^t(g)=b(g^{-1}), \quad b(g)a(g^{-1}) +d(g)b(g^{-1})
.\ee
The field equations of the model (1) are given by
\be (R+\Pi(G))\lambda +\partial_- G G^{-1}=0,\ee
\be \bar\Lambda (R+\Pi(G))-D_+G G^{-1}=0\ee
and 
\be D_+\chi -\partial_- \bar\Xi +[\bar\Xi,\chi]=0.\ee
Here $\chi$ and $\bar\Xi$ are understood to be elements of $\cgt$ and
their coordinates in the basis $\ti T_a$ are related to $\lambda$ and 
$\bar\Lambda$ as follows
\be \chi^a \equiv -d(G)_{ba}\lambda^b, \quad 
\bar\Xi^a \equiv - d(G)_{ba}\bar\Lambda^b.\ee
Note, that the equation (10) can be interpreted as the (1,0) SUSY 
zero-curvature
condition. Indeed, it can be integrated to give
\be \chi=-\partial_-\ti H \ti H^{-1}, \quad \bar\Xi = -D_+\ti H \ti H^{-1},\ee
where 
\be \ti H=\ti h+\theta \ti\eta \ti h, \quad \ti h\in \ti G, 
\quad \ti\eta\in\cgt.\ee
Thus we have obtained the (1,0) version 
 of the standard fact (cf. \cite{KS2}) that
every solution $G(\si,\tau,\theta)$ of the model (1) gives a string 
configuration $\ti H(\si,\tau,\theta)$ propagating in the dual target $\ti G$.

As in the bosonic case \cite{KS2}, we shall prove  that the configuration
$\ti G(\tau,\si,\theta)$, defined by
\be G(\tau,\si,\theta)\ti H(\tau,\si,\theta)=\ti G(\tau,\si,\theta) 
H(\tau,\si,\theta), \quad \ti G(\tau,\si,\theta)\in \ti G, 
\quad H(\tau,\si,\theta)\in G,\ee
 is a solution of the dual (1,0)  $\si$-model . The action of the latter
 is the same as of (1), except all quantities bear tilde and $R_{ab}$
  is replaced
 by its inverse matrix; explicitly:
 \be \ti S={1\over 2}\int d\si d\tau d\theta 
\{\ti{\bar\Lambda}_a((R^{-1})^{ab}+\ti\Pi(\ti G)^{ab})\ti\lambda_b + 
\ti{\bar\Lambda}_a (\partial_-\ti G \ti G^{-1})^a
-(D_+\ti G \ti G^{-1})^a\ti\lambda_a\}.\ee
To prove the statement, note that, using (11) and (12),
the field equations (8) and (9) can be 
rewritten
as
\be \la D_+L L^{-1},R^-_a\ra=0, \quad R^-_a\equiv R_{ba} T^b-\ti T_a,\ee
\be \la \partial_- L L^{-1},R^+_a\ra =0,\quad R^+_a\equiv 
 R_{ab}T^b +\ti T_a,\ee
where 
\be L(\tau,\si,\theta)=G(\tau,\si,\theta)\ti H(\tau,\si,\theta).\ee
In the same way, the field equations of the dual model are
 \be \la D_+\ti L \ti L^{-1},(R^{-1})^{ba} \ti T_b- T^a\ra=0,\ee
\be \la \partial_- \ti L \ti L^{-1}, (R^{-1})^{ab}\ti T_b +T^a\ra=0,\ee
where $\ti L(\tau,\si,\theta)=\ti G(\tau,\si,\theta) H(\tau,\si,\theta)$.
Thus, under the identification $L=\ti L$, the equations (16) and (17)
are the same as (19) and (20). This finishes the proof\footnote{It is,
 perhaps, worth
adding a comment concerning the global topology of the world sheet.
In general, $\ti G(\tau,\si,\theta)$ may turn out not to be single-valued and,
 thus, not to correspond to a closed string solution. If we restrict
 the space of solutions of the model (1) in such a way that 
 $\ti H(\tau,\si,\th)$
 is single-valued (and the same thing we do with $H(\tau,\si,\th)$ in the
  dual model (15)) then both $G(\tau,\si,\th)$ and $\ti G(\tau,\si,\th)$
  will be single valued.  Actually, we must to impose this restriction in order
  to have the "off shell" duality, see page 6.}.

3. The equations (16) and (17) are duality invariant ($L=\ti L$). 
It is therefore 
natural to ask, if there is a duality invariant formulation of the dynamics
of both models (1) and (15). The answer to this question is positive; 
we can 
construct this action starting directly from the expression (1). We shall
work in components; first note
\be D_+G G^{-1}=\g +\theta (\partial_+ g g^{-1}+\g\g),\ee
\be \partial_-G G^{-1}=\partial_- g g^{-1}+
\theta (\partial_-\g +[\g ,\partial_-g g^{-1}]),\ee
\be \bar\Xi\equiv\bk +\theta \bx, \qquad \chi \equiv x+\th \k.\ee
Of course, the components of $\bar\Xi$ and $\chi$ are also in $\cgt$;
$\bx, x$ are real bosons and $\bk,\k$ are Majorana-Weyl fermions.
We perform the integration over $\th$ to arrive at
\bea S={1\over 2}\int d\si 
d\tau \{\la \bx,g^{-1}T^a g\ra\la g^{-1}R^+_a g,x\ra-\la\bx,g^{-1}
\d_- g\ra+\la g^{-1}\d_+ g-g^{-1}\g\g g,x\ra\cr
-\la\bk,g^{-1}T^a g\ra\la g^{-1}R^+_a g,\k\ra +\la\bk,g^{-1}\d_-\g g\ra-
\la g^{-1}\g g,\k\ra~~~~~~~~~\cr -\la\bk ,
g^{-1}[T^a,\g ]g\ra\la g^{-1}R^+_a g,x\ra
-\la \bk, g^{-1} T^a g\ra\la g^{-1}[R^+_a,\g] g, x\ra \}.~~~\eea
Varying with respect to $\k$, we obtain
\be \g+\la \bk,g^{-1}R^-_a g\ra T^a=0,\ee
or
\be \la \psi,R^-_a\ra=0, \quad\psi\equiv \g -g\bk g^{-1}.\ee 
In the course of the derivation, the fermions $\g$, $\bk$ and $\psi$ will
be understood to fulfil the constraints (25,26). Now redefining
\be \bx =\bx'-\la \{\bk, g^{-1}\g g\}, T^a\ra \ti T_a,\ee
 we have
 \bea S={1\over 2}
 \int d\si d\tau \{\la\bx',g^{-1}T^a g\ra\la g^{-1}R^+_a g, x\ra
 -\la \bx',g^{-1}\d_- g\ra\cr
 +\la g\bk g^{-1},\d_-\g +[\g,\d_-g g^{-1}]\ra +
 \la g^{-1}\d_+ g +g^{-1}\psi\psi g ,x\ra\}.\eea
 Set
 \be x^a=\rho^a+\pi^a \equiv 
 \rho^a +\la \pas l l^{-1} +{1\over 2}\psi\psi,gT^a g^{-1}\ra, \ee
 \be \bx^a =\rho^a-\pi^a \equiv \rho^a -
 \la \pas l l^{-1} +{1\over 2}\psi\psi,gT^a g^{-1}\ra,\ee
 \be   l=g\ti h, \quad \ti h\in \ti G \ee
 and eliminate $\rho$:
 \bea S={1\over 2}\int d\si d\tau \{2\la \pas \ti h \ti h^{-1}, g^{-1} \d_- g
 \ra +\la \bk\bk , g^{-1} \d_- g\ra +\la g\bk g^{-1}, \d_- \g\ra\cr
 -\la \pas l l^{-1} +{1\over 2}\psi\psi,(A-Id)
(\pas l l^{-1}+{1\over 2}\psi\psi)\ra\} .\eea
Here $A$ is a linear idempotent self-adjoint map from the Lie algebra
$\cd$ of the double into itself. It has two equally degenerate
eigenvalues, $+1$ and $-1$, and the corresponding eigenspaces are linear
envelopes of
$R^+_a$ and $R^-_a$, respectively.
 $Id$ is the identity map from $\cd$ to $\cd$.

The transformation (29,30) deserves some explanation. We have 
traded the field $\pi\in\cgt$ for a group-valued field $\ti h\in\ti G $.
Indeed, it is easy to see that
\be \pi^a=(\pas \ti h \ti h^{-1})^a +
{1\over 2} \la \psi\psi, g T^a g^{-1}\ra.\ee
In order to write the duality invariant action, we have to require that
$\ti h(\tau,\si)$ is a single-valued function of $\si$ (otherwise
the string configuration $\ti g(\tau,\si)$ in the dual target would not be
closed). This sets a constraint
of unit monodromy of  the quantity (cf. footnote 1)
\be\pi -
{1\over 2}\la \psi\psi, g T^a g^{-1}\ra \ti T_a \in \cgt.\ee
 This is the analogue
of what happened in the bosonic case  \cite {KS2,KS3} and the resolution
of the issue is the same: we have to constrain the model (1) by allowing
only those of its solutions, which fulfil the constraint of the unit monodromy
of the quantity (34). The same must be obviously true also in the
dual case (15), where the quantity dual to (34) is constructed in the
exactly corresponding way. Only after imposing both constraints, the PL
T-duality between (1) and (15) takes place.

Consider the Polyakov-Wiegmann formula \cite{PW}
\be I(g\ti h)=I(g) +I(\ti h) +{1\over 4\pi} \la \pas \ti h \ti h^{-1},
g^{-1} \pat g\ra,\ee
where
\be I(l)\equiv {1\over 8\pi}\{\int d\si d\tau 
\la \pas l l^{-1},\pat l l^{-1}\ra+
{1\over 6}d^{-1}\la dl~l^{-1},[dl~l^{-1},
dl~l^{-1}]\ra\}.\ee
Using  PW formula (35), Eq. (25) and the fact that both algebras 
$\cg$ and $\cgt$
are isotropic, we arrive at

$$ S[l(\tau,\si),\psi(\tau,\si)]=$$
\bea={1\over 8\pi}\int  \biggl\{\la \pas l~l^{-1},\partial_- l~l^{-1}\ra +
{1\over 6}d^{-1}\la dl~l^{-1},[dl~l^{-1},
dl~l^{-1}]\ra -{1\over 2}\la \psi,\partial_- \psi\ra 
\cr -\la \pas l l^{-1} +{1\over 2}\psi\psi,(A-Id)
(\pas l l^{-1}+{1\over 2}\psi\psi)\ra \biggl\}.\eea
Of course, $\psi\in\cd$ is thought to fulfil $\la\psi,R^-_a\ra=0$.

The action (37) is the duality invariant first-order action, that we
have been looking for.  Since it
was directly derived from the $\si$-model action (1), it is
equivalent to it (with the constraint of the unit monodromy).
In particular, it must give the same
equations of motion. As a simple check of correctness of our calculation
we can derive those equations directly from (37). In doing
that one has to fix conveniently  a small gauge symmetry $l(\tau,\si)\to
l(\tau,\si)l_0(\tau)$. The field equations then read
\be \la \psi, R^-_a\ra=0, \quad \la \partial_+ l l^{-1} +
\psi\psi, R^-_a\ra=0,\ee
\be \la \d_- l l^{-1}, R^+_a\ra =0, 
\quad \la \d_- \psi +[\psi, \d_- l l^{-1}], R^+_a\ra=0.\ee
If we combine the fields $l$ and $\psi$ in a single group-valued superfield
\be L=l+\th\psi l,\ee
we can rewrite the component equations (38) and (39) as 
\be \la D_+L L^{-1}, R^-_a\ra= 0, \quad \la \d_- L L^{-1}, R^+_a\ra =0.\ee
Those are precisely the field equations (16) and (17). 

\newpage

4. Consider now the (1,1) SUSY extension of the PL dualizable $\sigma$-model
on the target $G$:

\be S={1\over 2}\int d\si d\tau d\theta^+ d\th^- 
\{\Lambda^a_+ (R_{ab}+\Pi(G)_{ab})\Lambda^b_- +
\Lambda^a_+ (D_-G G^{-1})_a
-(D_+ G G^{-1})_a\Lambda^a_-\}.\ee
Now we have two Grassmann world-sheet coordinates
$\theta^{\pm}$,  the superpartners of $\xi^{\pm}={1\over 2}(\tau\pm\si)$,
respectively.  Hence
\be D_{\pm}=\partial_{\theta^{\pm}}+\theta^{\pm}\partial_{\pm}.\ee
The superfield $G$ is given in components as
\be G=(1+\theta^+ \gamma^+ +\th^- \g^- +\th^+\th^-(K +{1\over 2}[\g^-,\g^+])) g,
 \quad g\in G, \quad \g^{\pm},K\in\cg;\ee
$\g^{\pm}$ are  components of the Majorana fermion, $K$ is a real auxiliary
bosonic field.
The odd auxiliary superfields $\Lambda^a_{\pm}$
can be written in the standard way as expansions in $\th$-s with generic
coefficients.

The field equations of the model (42) are given by
\be (R+\Pi(G))\Lambda_- +D_- G G^{-1}=0,\ee
\be \Lambda_+ (R+\Pi(G))-D_+G G^{-1}=0\ee
and 
\be D_+\Xi_- + D_- \Xi_+ +\{\Xi_+,\Xi_-\}=0.\ee
Here $\Xi_{\pm}$ are understood to be elements of $\cgt$ and
their coordinates in the basis $\ti T_a$ are related to $\Lambda_{\pm}$
 as follows
\be 
\Xi^a_{\pm} \equiv - d(G)_{ba}\Lambda^b_{\pm}.\ee
Note, that the equation (47) can be interpreted as the (1,1) SUSY 
zero-curvature
condition. Indeed, it can be integrated to give
\be \Xi_{\pm} = -D_{\pm}\ti H \ti H^{-1},\ee
where 
\be \ti H=
(1+\theta^+ \ti\eta^+ +\th^- \ti\eta^- +\th^+\th^-(\ti M +
{1\over 2}[\ti\eta^-,\ti\eta^+]))\ti h, 
\quad \ti h\in \ti G, 
\quad \ti\eta^{\pm}, \ti M\in\cgt.\ee
If we set
\be L(\tau,\si,\th^{\pm})=G(\tau,\si,\th^{\pm})\ti H(\tau,\si,\th^{\pm}),\ee
then the field equations (45) and (46) can be written as (cf. (16),(17)):
\be \la D_{\pm} L L^{-1},R^{\mp}_a\ra=0.\ee
The dual (1,1) $\si$-model ( the interested reader can easily
write its action by 
combining (1),(15) and (42)) has the identical equations of motions
and we can look for the duality invariant formulation of the common
dynamics of both models which would use only $L$ as a configuration. 
We can proceed exactly as in the (1,0) case by eliminating (all but one
component of the)
auxiliary fields $\Lambda_{\pm}$. The resulting duality invariant action
is given by

\bea S[l(\tau,\si),\psi^{\pm}(\tau,\si)]=
{1\over 8\pi}\int  \biggl\{\la \pas l~l^{-1},\pat l~l^{-1}\ra +
{1\over 6}d^{-1}\la dl~l^{-1},[dl~l^{-1},
dl~l^{-1}]\ra\cr -{1\over 2}\la \psi^+,\partial_- \psi^+\ra 
+{1\over 2}\la \psi^-,\partial_+ \psi^-\ra 
  +{1\over 4}\la \psi^+\psi^+,\psi^+\psi^+\ra 
-{1\over 4}\la \psi^-\psi^-,\psi^-\psi^-\ra\cr
+\la \pas l l^{-1}, \psi^+\psi^+ + \psi^-\psi^-\ra -
{1\over 4}\la \{\psi^-,\psi^+\}, A\{\psi^-,\psi^+\}\ra\cr
-\la \pas l l^{-1} +{1\over 2}\psi^+\psi^+ -{1\over 2}\psi^-\psi^-,A
(\pas l l^{-1}+{1\over 2}\psi^+\psi^+ -{1\over 2}\psi^-\psi^-)\ra \biggl\}.\eea
Here $l\in D$ and  $\psi^{\pm}$ are components
of the Majorana spinor with values in $\cd$, 
such that $\la \psi^{\pm},R^{\mp}_a\ra=0$.

This action (53) is our final result. Note that it has the correct (1,0)
limit (if we set $\psi^-$ equal $0$) and correct (0,0) limit (\cite{KS3})
if we set both spinors $\psi^{\pm}$ to zero. We may also find easily the
field equations following from (53). They are

\be \la \psi^{\pm},R^{\mp}_a\ra=0,\quad \la \d_{\pm}l l^{-1} +
\psi^{\pm}\psi^{\pm},R^{\mp}_a\ra=0,\ee
\be \la \d_{\pm}\psi^{\mp}+[\psi^{\mp},\d_{\pm}l l^{-1} +{1\over 2} \psi^{\pm}
\psi^{\pm}] \pm {1\over 2}[A\{\psi^-,\psi^+\},\psi^{\pm}],R^{\mp}_a\ra=0.\ee
If we combine the fields $l$ and $\psi^{\pm}$ in a single group-valued 
superfield
\be L=(1+\th^+\psi^+ + \th^- \psi^- +\th^+\th^- 
(F +{1\over 2}[\psi^-\psi^+])) l,\ee
($F\in\cd$ is an auxiliary field) we can rewrite the component equations (54)
 and (55) as 
\be \la D_{\pm}L L^{-1}, R^{\mp}_a\ra= 0.\ee
Those are precisely the field equations (52).

\vskip2pc

I thank P. \v Severa and K. Sfetsos for discussions and correspondence.

 
\end{document}